# Dynamic Polarizability of Low-Dimensional Excitons

Thomas Garm Pedersen

*Department of Materials and Production, Aalborg University, DK-9220 Aalborg Øst, Denmark*
*Email: tgp@mp.aau.dk*

Excitons in low-dimensional materials behave mathematically as confined hydrogen atoms. An appealing unified description of confinement in quantum wells or wires etc. is found by restricting space to a fractional dimension $1 < D \leq 3$ serving as an adjustable parameter. We compute the dynamic polarizability of $D$-dimensional excitons in terms of discrete and continuum oscillator strengths. Analyzing exact sum rules, we show that continuum contributions are increasingly important in low dimensions. The dynamical responses of excitons in various dimensions are compared. Finally, an exact and compact closed-form expression for the dynamic polarizability is found. This completely general formula takes $D$ as input and provides exact results for arbitrary frequency.

## 1. Introduction

Low-dimensional hydrogen has been extensively applied as a model of excitons in low-dimensional materials, in particular, two-dimensional semiconductors and quantum wells [1-3]. This Wannier-Mott approach exploits the mathematical similarities between electron-proton interactions in atoms and electron-hole interactions in semiconductors. If screening can be assumed local, i.e. spatially constant, and an effective mass approximation is adopted, the connection is particularly simple. Thus, the bare hydrogen Bohr radius $a_0$ and Hartree energy Ha are replaced by effective quantities $a_0^* = (\varepsilon m_0 / \mu) a_0$ and $\text{Ha}^* = (\mu / m_0 \varepsilon^2) \text{Ha}$, respectively, where $\varepsilon$ is the dielectric constant, $m_0$ the free-electron mass, and $\mu = m_e m_h / (m_e + m_h)$ the reduced effective mass given in terms of effective electron and hole masses $m_e$ and $m_h$. It follows that hydrogen results carry over to semiconductors provided a simple scaling of distance and energy units is applied. While such models are often quantitatively correct in heterostructures, excitons in atomically thin two-dimensional materials experience pronounced nonlocal screening leading to non-hydrogenic Rydberg series [4,5]. Nevertheless, hydrogen-like states are frequently applied as a starting point for models of these materials as well [6]. In addition, quasi-two-dimensional hydrogen-like excitons emerge in extremely anisotropic bulk semiconductors, such as multilayer transition-metal dichalcogenides [7].

An atomically thin semiconductor is essentially two-dimensional in the sense that motion in the third dimension is effectively frozen by confinement dominating over Coulomb effects for the dynamics in this direction. In contrast, semiconductor quantum wells may have a thickness comparable to the effective Bohr radius and excitons extend into the third dimension. A simple and intuitive picture of such delocalization is through the concept of fractional dimensions [8-13]. Hence, the effective dimension $D$ of the exciton transitions from two to three as the quantum well widens. Similarly, excitons in quantum wires or nanotubes effectively behave as hydrogenic states in dimensions $1 < D < 2$. This picture has been extremely useful in deriving simple and closed form expressions for the interband absorption spectrum of excitons [8,9]. In the Wannier model, interband transition strength is determined by the exciton wavefunction of the final state evaluated at the coordinate origin, which physically corresponds to the amplitude for coinciding electron and hole locations. Since only *s*-states are finite at the origin, only these contribute to interband absorption [8,9].

Confined excitons perturbed by external fields may similarly be described by perturbation theory applied in fractional dimensional space. The Stark shift in static electric fields has been analyzed both at second and higher orders [11,13]. This procedure provides static (hyper-) polarizabilities and field-induced ionization rates via resummation. Such polarizabilities and ionization rates have been successfully compared to experiments in two-dimensional semiconductors [14]. It is therefore highly desirable to extend the analysis to time-dependent external fields. This would allow for an analysis of field-induced excitation of excitons, as well as AC ionization and Stark effects that have recently been studied experimentally in transition-metal dichalcogenides [15,16]. The response to dynamic electric fields is significantly more complicated, however. The required tools from time-dependent perturbation theory are less amenable to analytical results than time-independent ones, for which the Dalgarno-Lewis approach [17] typically provides analytic closed-form results. Fully numerical results within this approach are still possible, though [18]. In the present work, we show how traditional sum-over-states perturbation theory is a highly useful alternative allowing us to find exact dynamical polarizabilities of low-dimensional excitons and even closed-form results for arbitrary dimensions including the especially important two-dimensional case.

## 2. Dynamic Polarizability

We consider the response of low-dimensional Wannier-Mott excitons to monochromatic electric fields. The aim is to find the frequency ($\omega$) dependent polarizability $\alpha_D(\omega)$ for *D*-dimensional excitons with *D* an unspecified parameter that may take non-integer values. This task requires exciton oscillator strengths, for which we establish several exact sum rules. Moreover, they lead to a simple closed-form expression for the dynamic polarizability of *D*-dimensional excitons. As mentioned above, obtaining analytical results for the problem at hand seems to be too complicated for the time-dependent Dalgarno-Lewis method and, so, all results have been derived using a classical sum-over-states approach. A challenge with this approach is that the continuum of ionized excited states must be handled with special care. Hence, the frequency dependent polarizability of the 1*s* ground state is

$$\alpha_D(\omega) = \sum_n \frac{g_D(n)}{E_n^2 - \omega^2} + \int_0^\infty \frac{g_D'(k)}{E_k^2 - \omega^2} dk, \tag{1}$$

where $g_D(n)$ is the oscillator strength for discrete transitions while $g'_D(k)$ is the oscillator strength distribution for the continuum. Throughout, we use natural exciton units taking $a_0^*$ and Ha$^*$ as units of distance and energy, respectively. Polarizabilities are reported as dimensionless quantities but may be converted into ordinary units through multiplication by a factor $4\pi\varepsilon_0 a_0^3(\text{Ha}/\text{Ha}^*)^2$. In $D$-dimensional hydrogen, energies are $-k_n^2/2$ with $k_n = 2/(2n+D-3)$ and $n$ a positive integer. Hence, denoting the ground state wave number as $k_D = 2/(D-1)$, we find transition energies $E_n = (k_D^2 - k_n^2)/2$ for discrete transitions and $E_k = (k_D^2 + k^2)/2$ for the continuum. The optical field is assumed to point along an un-confined direction ($x$) and, so, only $p$-states couple to the ground state. Consequently, the oscillator strength is $g_D(n) = 2E_n |\langle \varphi_{np}|x|\varphi_{1s}\rangle|^2$ with $n \geq 2$ and the dipole matrix element between the 1$s$ ground state and $np$ excited state given by

$$\langle \varphi_{np}|x|\varphi_{1s}\rangle = \frac{1}{\sqrt{D}} \int_0^\infty R_{np}(r) R_{1s}(r) r^D dr . \tag{2}$$

In this expression, the prefactor follows from the angular integration and the required radial eigenfunctions are $R_{1s}(r) = (2k_D)^{D/2} e^{-k_D r}/\sqrt{\Gamma(D)}$ and

$$R_{np}(r) = (2k_n)^{(D+1)/2} e^{-k_n r} k_n r \sqrt{\frac{(n-2)!}{\Gamma(D-1+n)}} L_{n-2}^D(2k_n r), \tag{3}$$

where $L_n^m$ is an associated Laguerre polynomial. The required integral can be carried out analytically with the result for the oscillator strength

$$g_D(n) = \frac{(n-1)^{2n-4}(D-1)^{2+D}(2n-3+D)^{1+D}\Gamma(n-2+D)}{(n-2+D)^{2(n+D-1)}(n-1)!\Gamma(1+D)}. \tag{4}$$

It is readily verified that well-know integer-dimensional cases are reproduced as special cases of this more general result, i.e. [19]

$$g_3(n) = \frac{256(n-1)^{2n-4} n^5}{3(n+1)^{2n+4}},$$
$$g_2(n) = \frac{4(n-1)^{2n-4}(n-\frac{1}{2})^3}{n^{2n+2}}. \tag{5}$$

Next, we need to handle transitions to the continuum of ionized states with positive energy $k^2/2$. These states are similar to Eq.(3) but with the substitution $k_n \to ik$ and the hypergeometric function ${}_1F_1\left[\frac{D+1}{2}+\frac{i}{k}, D+1, 2ikr\right]$ replacing the Laguerre polynomial. We use the important result

$$\int_0^\infty e^{-k_D r - ikr} (kr)^{D+1} {}_1F_1\left[\frac{D+1}{2} + \frac{i}{k}, D+1, 2ikr\right] dr$$
$$= \Gamma(D+1) \frac{(D+1)k_D - 2}{k^2} \left(\frac{k^2}{k^2 + k_D^2}\right)^{\frac{D+3}{2}} \exp\left\{-\frac{2}{k} \tan^{-1} k/k_D\right\}. \tag{6}$$

In turn, the oscillator strength distribution due to the continuum becomes

$$g'_D(k) = \frac{4^{D+1}}{\pi \Gamma(D+1)k} \left(\frac{kk_D}{k_D^2 + k^2}\right)^{D+2} \exp\left\{\frac{\pi - 4\tan^{-1} k/k_D}{k}\right\} \left|\Gamma\left(\frac{D+1}{2} - \frac{i}{k}\right)\right|^2. \tag{7}$$

Again, it may be verified that this general result conforms to special integer-dimensional cases [19]

$$g'_3(k) = \frac{2^8 k}{3(1+k^2)^4} \frac{\exp\left\{-\frac{4}{k}\tan^{-1} k\right\}}{1 - \exp\{-2\pi/k\}},$$
$$g'_2(k) = \frac{2^8 k}{(4+k^2)^3} \frac{\exp\left\{-\frac{4}{k}\tan^{-1} k/2\right\}}{1 + \exp\{-2\pi/k\}}. \tag{8}$$

It may also be noted, however, that the general case with arbitrary $D$ cannot be further simplified due to the presence of gamma functions with complex arguments. Nevertheless, several important exact sum rules are readily established in terms of matrix elements of the ground state. To this end, we consider the $p$'th transition energy moment of the sum over oscillator strengths

$$S_p = \sum_n g_D(n) E_n^p + \int_0^\infty g'_D(k) E_k^p dk. \tag{9}$$

These are evaluated in terms of matrix elements of the ground state only. In this manner, generalizing the results of Jackiw [20] to arbitrary dimensions, we find

$$S_{-2} = \frac{(D-1)^4(D+1)(2D+3)}{128}, \; S_{-1} = \frac{(D-1)^2(D+1)}{8},$$
$$S_0 = 1, \; S_1 = \frac{8}{(D-1)^2 D}, \; S_2 = \frac{64}{(D-1)^4(D-2)D}. \tag{10}$$

These have all been verified numerically to high precision. Note that $S_2$ is finite only for $D > 2$. The negative second moment is the static polarizability [11,13] $\alpha_D(0) = S_{-2}$, while $S_0 = 1$ is the Thomas-Reiche-Kuhn [21,22] sum rule.

As an interesting application, we show in Fig. 1 the contribution to $S_0$ from discrete transitions solely, i.e. retaining only the discrete sum in Eq.(9). Thus, the graph provides a measure of the importance of continuum states in the dynamic response. It is clearly seen that, in low dimensions, the discrete oscillator strength contribution is minor. In fact, at $D=2$ the fraction is 28%, while at $D=1.71$, as appropriate for carbon nanotubes [11], it is as low as 17%. Analytically, it can be shown that $\sum_n g_D(n) \approx 4\zeta(3)(D-1)^3 \approx 4.81(D-1)^3$ as $D \to 1$, i.e. approaching the strict one-dimensional limit. Conversely, in high dimensions, the discrete contribution dominates, as seen in the inset. Similarly, the static polarizability due to discrete transitions for quasi-one-dimensional excitons is $\alpha_D(0)_{\text{discrete}} \approx \zeta(3)(D-1)^7$, severely underestimating the true result. It is, therefore, clear that transitions to the continuum are indispensable for quantitatively correct results.

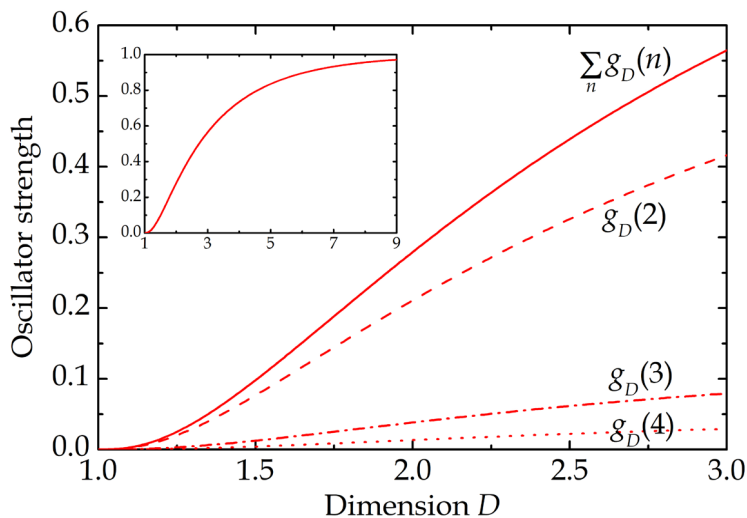

Figure 1. Individual and summed oscillator strengths from discrete transitions, i.e. omitting the continuum. The inset shows the behavior of the sum for large dimensions, asymptotically approaching unity.

## 3. Results

We now turn to the dynamic polarizability of excitons in various dimensions. Based on Eqs.(4) and (7) for the discrete and continuous contributions, respectively, the general sum-over-states expression Eq.(1) is readily evaluated by numerical summation (in practice summing to $n=500$ and integrating using quadrature). Since the poles of the spectra are known analytically, however, we are able to derive a simple closed-form expression for the entire spectrum below. This is accomplished by taking inspiration from the remarkable closed-form result for three dimensions found by Gavrila [23] that can be shown to agree with other differently formulated but equivalent results [24,25]

$$\alpha_3(\omega) = -\frac{1}{2\omega^2} + \frac{128(1-2\omega)^2}{\omega^2(\sqrt{1-2\omega}+1)^8(2\sqrt{1-2\omega}-1)} \times \quad (11)$$

$$_2F_1\left[4, 2-\tfrac{1}{\sqrt{1-2\omega}}, 3-\tfrac{1}{\sqrt{1-2\omega}}, \left(\tfrac{\sqrt{1-2\omega}-1}{\sqrt{1-2\omega}+1}\right)^2\right] + (\omega \to -\omega).$$

The elegant and compact result Eq.(11) is readily demonstrated to agree with the numerical sum-over-states Eq.(1) for $D=3$ as seen in Fig. 2. In all spectral plots, a finite line width is included by the substitution $\omega \to \omega + i\gamma$ with broadening $\gamma = 0.005 k_D$. The correctness of the analytical result Eq.(11) follows from the pole structure of the hypergeometric function. Thus, $_2F_1[a,b,c,z]$ has poles whenever $c = m$ with $m$ a negative integer [26]. In the three-dimensional case, taking $m = 1-n$ with $n \geq 2$, this translates into resonances at $\omega = (1-\tfrac{1}{n^2})/2$ as required. In the $D$-dimensional case, this suggests a general structure

$$\alpha_D(\omega) = -\frac{1}{2\omega^2} + f_D(\omega) \,_2F_1\left[1+D, \tfrac{1+D}{2}-\tfrac{1}{\sqrt{k_D^2-2\omega}}, \tfrac{3+D}{2}-\tfrac{1}{\sqrt{k_D^2-2\omega}}, \left(\tfrac{\sqrt{k_D^2-2\omega}-k_D}{\sqrt{k_D^2-2\omega}+k_D}\right)^2\right] + (\omega \to -\omega). \quad (12)$$

In particular, the $c$ argument is fixed by the $D$-dimensional resonances. It is then relatively straightforward to match the poles to the oscillator strengths given by Eq.(4). In this manner,

$$f_D(\omega) = \frac{2^{1+2D} k_D^{3+D} (k_D^2 - 2\omega)^{(1+D)/2}}{\omega^2 [k_D + \sqrt{k_D^2 - 2\omega}]^{2(1+D)} [(1+k_D)\sqrt{k_D^2-2\omega} - k_D]}. \quad (13)$$

This exact and compact expression for the dynamic polarizability is the most important result of the present work. Given the complexity of the oscillator strengths Eqs.(4) and (7), it is remarkable that such a simple closed form exists. It is found to capture all poles of the numerical spectra. Furthermore, the expression behaves appropriately in both low- and high-frequency limits. Thus, at low frequency

$$\alpha_D(\omega) = \frac{(D-1)^4(D+1)(2D+3)}{128} + \frac{(D-1)^8(D+1)(137+220D+123D^2+24D^3)}{98304}\omega^2 + O(\omega^4). \quad (14)$$

The first term of this expansion is precisely the known static polarizability [11,13], while the second term is new. As a particularly important application, we find for $D=2$

$$\alpha_2(\omega) = -\frac{1}{2\omega^2} + \frac{1024(4-2\omega)^{3/2}}{\omega^2(\sqrt{4-2\omega}+2)^6(3\sqrt{4-2\omega}-2)} \times \quad (15)$$

$$_2F_1\left[3, \tfrac{3}{2}-\tfrac{1}{\sqrt{4-2\omega}}, \tfrac{5}{2}-\tfrac{1}{\sqrt{4-2\omega}}, \left(\tfrac{\sqrt{4-2\omega}-2}{\sqrt{4-2\omega}+2}\right)^2\right] + (\omega \to -\omega).$$

As seen in Figs. 2 and 3, this compact expression is in perfect agreement with the numerical evaluation. Moreover, we find the analytical low-frequency limit $\alpha_2(\omega) = 21/128 + \omega^2 1261/32768 + O(\omega^4)$ in agreement with known results [11,13]. The high-frequency behavior is $\sim -1/\omega^2$ as required by the Thomas-Reiche-Kuhn sum rule. Comparing spectra for various dimensions in Fig. 2, several trends are noted. With the scaling applied in the plot, all four cases look quite similar. Apart from the obvious scaling with energy, the ratio of oscillator strengths for the dominant transitions is only weakly dependent on D, i.e. $g_2(3)/g_2(2) \approx 0.18$ while $g_3(3)/g_3(2) \approx 0.19$. The overall scale of the spectra decreases, however, as dimension is reduced, as evidenced by the oscillator strengths in Fig. 1. This is in line with the static polarizability varying as $\alpha_D(0) \sim (D-1)^4$ as $D \to 1$.

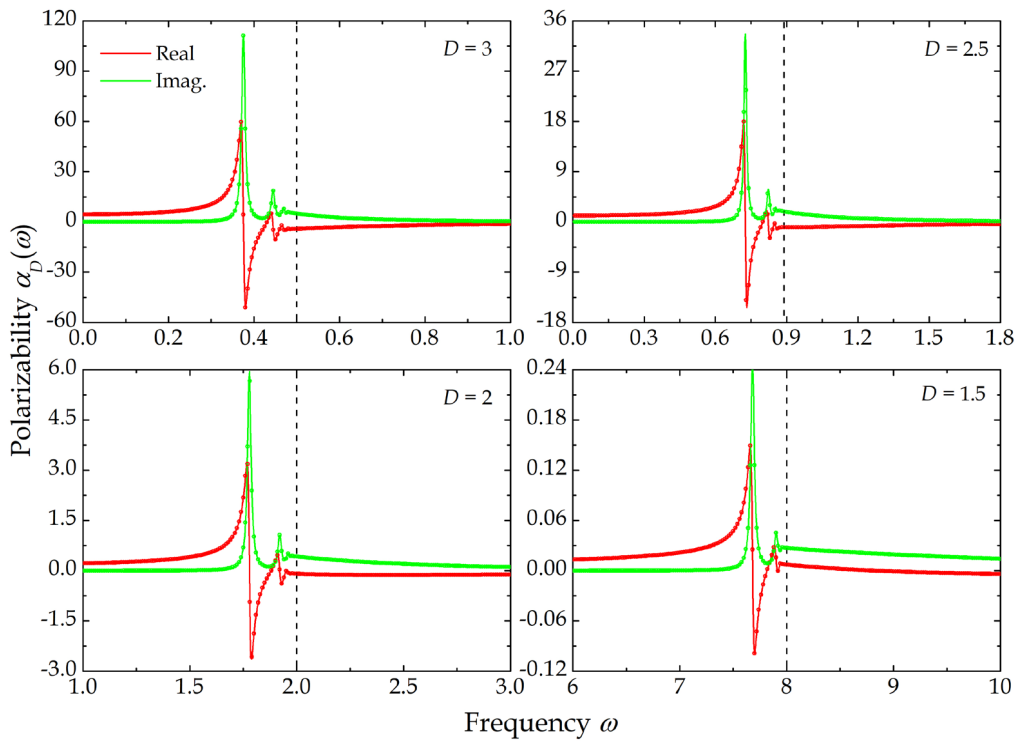

Figure 2. Real and imaginary parts of the polarizability for four different dimensions. The vertical dashed lines indicate the onset of the continuum $k_D^2/2$ and the circles are analytical results.

An advantage of the sum-over-states expression Eq.(1), as compared with closed-form results such as Eqs.(11), (12) or (15), is that contributions from discrete and continuum transitions are easily separated. This provides a measure of the error introduced by ignoring continua, as is commonly done in fully numerical approaches. In Fig. 3, we split the response for two-dimensional materials into discrete and continuum parts. In addition, we again illustrate how their sum agrees with the analytical result. It is clearly seen that the absorptive imaginary part is well approximated by discrete contributions only, as long as the frequency is below the ionization threshold $k_D^2/2$, which equals 2 in this case. The real part, however, is severely affected by continuum contributions. This is most clearly seen in the vicinity of the ionization threshold. Yet, the low-frequency response is also greatly underestimated using only discrete contributions. We find $\alpha_2(0)_{\text{discrete}} \approx 0.0848$, to be compared with the full result $\alpha_2(0) = 21/128 \approx 0.164$. Thus, the error introduced by omitting the continuum is close to 50%. Near the ionization threshold, the error is even greater.

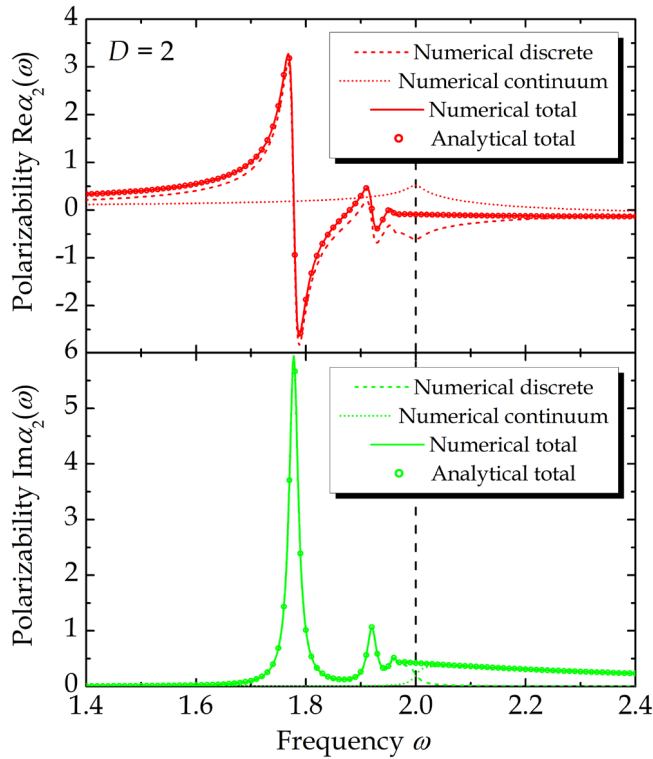

Figure 3. Real and imaginary parts of the polarizability of two-dimensional excitons. Discrete, continuum, and total contributions are shown as dashed, dotted and solid lines, respectively, while circles are analytical results.

## 4. Summary

In summary, we have computed the dynamic polarizability of Wannier-Mott excitons in arbitrary dimensions $D$. Exact $D$-dimensional oscillator strengths for both discrete and continuum transitions have been derived. Moreover, several exact sum rules have been established and applied to estimate the error inherent in omitting continuum contributions. The pole structure in the general case is analyzed and utilized to construct a simple and compact closed-form expression for the polarizability of $D$-dimensional excitons. An analysis of this exact result implies that the real part of the exciton polarizability is severely underestimated if only discrete-state transitions are considered.

## References

1. H. Haug and S.W. Koch, "Quantum Theory of the Optical and Electronic Properties of Semiconductors" (World Scientific, Singapore, 1993).

2. P. K. Basu, "Theory of optical processes in semiconductors" (Oxford University Press, Oxford, 1997).
3. T. Cheiwchanchamnangij and W. R. L. Lambrecht, Phys. Rev. B 85, 205302 (2012).
4. A. Chernikov, T. C. Berkelbach, H. M. Hill, A. Rigosi, Y. Li, O. B. Aslan, D. R. Reichman, M. S. Hybertsen, and T. F. Heinz, Phys. Rev. Lett. 113, 076802 (2014).
5. D. Van Tuan, M. Yang, and H. Dery, Phys. Rev. B 98, 125308 (2018).
6. T. Olsen, S. Latini, F. Rasmussen, and K. S. Thygesen, Phys. Rev. Lett. 116, 056401 (2016).
7. T. G. Pedersen, S. Latini, K. S. Thygesen, H. Mera, and B.K. Nikolic, New J. Phys. 18, 073043 (2016).
8. X-F. He, Phys. Rev. B 43, 2063 (1991).
9. P. Lefebvre, P. Christol, and H. Mathieu, Phys. Rev. B 48, 17308 (1993).
10. M. A. Lohe and A. Thilagam, J. Phys. A. Math. Gen. 37, 6181 (2004).
11. T. G. Pedersen, Solid State Commun. 141, 569 (2007).
12. T. F. Rønnow, T. G. Pedersen, and B. Partoens, Phys. Rev. B 85, 045412 (2012).
13. T. G. Pedersen, H. Mera, and B. K. Nikolić, Phys. Rev. A 93, 013409 (2016).
14. M. Massicotte, F. Vialla, P. Schmidt, M. Lundeberg, S. Latini, S. Haastrup, M. Danovich, D. Davydovskaya, K. Watanabe, T. Taniguchi, V. Fal'ko, K. Thygesen, T. G. Pedersen, and F. H. L. Koppens, Nat. Commun. 9, 1633 (2018).
15. J. Kim, X. Hong, C. Jin, S.-F. Shi, C.-Y. S. Chang, M.-H. Chiu, L.-J. Li, and F. Wang, Science 346, 1205 (2014).
16. E. J. Sie, J. W. McIver, Y.-H. Lee, L. Fu, J. Kong, and N. Gedik, Nature Mat. 14, 290 (2015).
17. A. Dalgarno and J. T. Lewis, Proc. R. Soc. Lond. A 233, 70 (1955).
18. J. C. G. Henriques, M. F. C. Martins Quintela, and N. M. R. Peres, J. Opt. Soc. Am. B 38, 2065 (2021).
19. X. L. Yang, S. H. Guo, F. T. Chan, K. W. Wong, and W. Y. Ching, Phys. Rev. A 43, 1186 (1991).
20. R. Jackiw, Phys. Rev. 157, 1220 (1967).
21. W. Kuhn, Z. Phys. 33, 408 (1925).
22. F. Reiche and W. Thomas, Z. Phys. 34, 510 (1925).
23. M. Gavrila, Phys. Rev. 163, 147 (1967).
24. S. I. Vetchinkin and S. V. Kchristenko, Chem. Phys. Lett. 1, 437 (1967).
25. C. M. Adhikari, V. Debierre, A. Matveev, N. Kolachevsky, and U. D. Jentschura, Phys. Rev. A 95, 022703 (2017).
26. I. S. Gradshteyn and I. M. Ryzhik "Table of integrals, series, and products", (Academic Press, Amsterdam, 2007).